\documentclass[pre, reprint, superscriptaddress, amsmath, amssymb, aps]{revtex4-2}

\usepackage[T1]{fontenc}
\usepackage[english]{babel}
\newcommand{\hyph}{\babelhyphen{hard}}

\usepackage{graphicx}
\usepackage{dcolumn}
\usepackage[protrusion=true,expansion=true]{microtype}
\usepackage[normalem]{ulem}
\usepackage{soul}
\usepackage{kotex}
\usepackage{hyperref}
\hypersetup{hidelinks}
\usepackage{cleveref}
\usepackage{comment}
\usepackage{xr}
\usepackage{mathtools}
\DeclarePairedDelimiter{\ceil}{\lceil}{\rceil}

\newcommand{\phct}{p_h^{c}}
\newcommand{\phsp}{p_h^{**}}
\newcommand{\Sent}{\sigma_{\text{ent}}}

\usepackage{bibunits}
\usepackage{etoolbox}
\makeatletter
\let\NAT@sort\z@
\let\NAT@cmprs\@ne
\renewcommand{\@bibunitname}{\jobname-\the\@bibunitauxcnt}
\AtBeginDocument{%
  \let\auto@bib\@empty
  \let\auto@bib@innerbib\@empty
}
\newcommand{\SetBibliographyUnit}[2]{%
  \def\@extra@b@citeb{#1}%
  \def\@extra@binfo{#1}%
  \renewcommand{\citenumfont}[1]{#2##1}%
  \renewcommand{\bibnumfmt}[1]{[#2##1]}%
}
\patchcmd{\NAT@bibsetnum}{LastBibItem}{LastBibItem\@extra@b@citeb}{}%
  {\PackageError{shyvoter}{Cannot separate bibliography label widths}{}}
\patchcmd{\endthebibliography}{LastBibItem}{LastBibItem\@extra@b@citeb}{}%
  {\PackageError{shyvoter}{Cannot separate final bibliography labels}{}}
\makeatother
\begin{document}


\title{A Shy Voter Model: Entropy Flux in Opinion Formation}

\author{Cook Hyun Kim}
\affiliation{CCSS, KI for Grid Modernization, Korea Institute of Energy Technology, Naju, Jeonnam\hyph{}Gwangju 58330, Korea}
\author{Sangjoon Park}
\affiliation{CCSS, KI for Grid Modernization, Korea Institute of Energy Technology, Naju, Jeonnam\hyph{}Gwangju 58330, Korea}
\author{B. Kahng}
\affiliation{CCSS, KI for Grid Modernization, Korea Institute of Energy Technology, Naju, Jeonnam\hyph{}Gwangju 58330, Korea}
\date{\today}

\begin{abstract}
Some voters conceal their opinions, causing a mismatch between visible and private opinions in polls. Conventional voter models do not account for such concealment. Here, we introduce a shy voter model that incorporates a hidden state in addition to two visible states, as an extension of the conventional majority\hyph{}rule voter model. The model exhibits continuous and hybrid transitions, yielding a rich phase diagram organized by the hidden fraction, i.e., the fraction of voters in the hidden state. For a large hidden fraction, the system undergoes a hybrid transition from the ordered state to the symmetric (disordered) state, whereas for a small hidden fraction, it exhibits a continuous transition, with a tricritical point separating the two regimes. Near criticality, both the order\hyph{}parameter variance and the susceptibility diverge. The hidden fraction also governs the entropy flux, a quantity that is not directly observable in real\hyph{}world situations. Thus, opinion concealment qualitatively reshapes opinion dynamics, while the present framework provides a basis for further extensions toward more realistic forms of social interactions.
\end{abstract}

\maketitle

\begin{bibunit}[apsrev4-2]
\SetBibliographyUnit{.main}{}


{\it Introduction:} 
Opinion poll estimates sometimes differ noticeably from election results. Researchers have proposed several explanations, including problems with sampling and weighting, inaccurate turnout estimates, and late changes in voter preferences~\cite{prosser2018polls,kennedy2018polls}. Opinion concealment is another proposed explanation: voters may withhold or misreport their views to avoid social disapproval or isolation~\cite{tourangeau2007sensitive,matthes2018spiral,noelle1974spiral,stout2015racial}.


Conventional voter models~\cite{castellano2009social,redner2019review,jedrzejewski2019review,starnini2026opinion,clifford1973competition,holley1975voter} and their majority\hyph{}rule variants~\cite{oliveira1992majority,krapivsky2003majority} assume that every voter expresses a visible opinion and participates in opinion transmission, even in extended formulations that incorporate nonlinear responses~\cite{castellano2009nonlinear,nyczka2012qvoter,min2017fragmentation,raducha2018coevolving,peralta2018nonlinear,min2019multilayer}, stochasticity~\cite{granovsky1995noisy,nyczka2012qvoter,peralta2018nonlinear,herrerias2019multistate}, committed agents~\cite{mobilia2003zealot,mobilia2007zealotry,xie2011committed}, group\hyph{}level effects~\cite{galam2002minority,krapivsky2003majority}, or private opinions~\cite{gastner2018concealed,gastner2019hypocrisy}. These models therefore do not address how opinion dynamics changes when some voters conceal their opinions and withdraw from opinion transmission. Describing this form of concealment requires an explicit hidden state that contributes no visible opinion to neighboring voters.

Here, we introduce a shy voter model in which one hidden state, denoted as $s=0$, is added to two visible states, $s=+1$ and $s=-1$. Consider a voter whose $k$ neighbors have a dominant visible opinion, say, $s=+1$. A voter who holds the opposite opinion may then avoid expressing it. Thus, the voter enters the hidden state with probability $p_h$ and adopts the majority visible opinion with probability $1-p_h$. When the visible opinions in the neighborhood are nearly evenly split, the voter may hesitate about which opinion to express and conceal it. To capture this, we introduce the visibility probability $p_v$. The voter conceals their opinion with probability $1-p_v$, or adopts $s=+1$ or $s=-1$ with probability $p_v/2$ each.
 
A small $p_v$ promotes concealment, whereas a large $p_v$ suppresses it. Thus, $p_v$ controls the steady\hyph{}state hidden fraction.

\begin{figure}[t]
\includegraphics[width=\columnwidth]{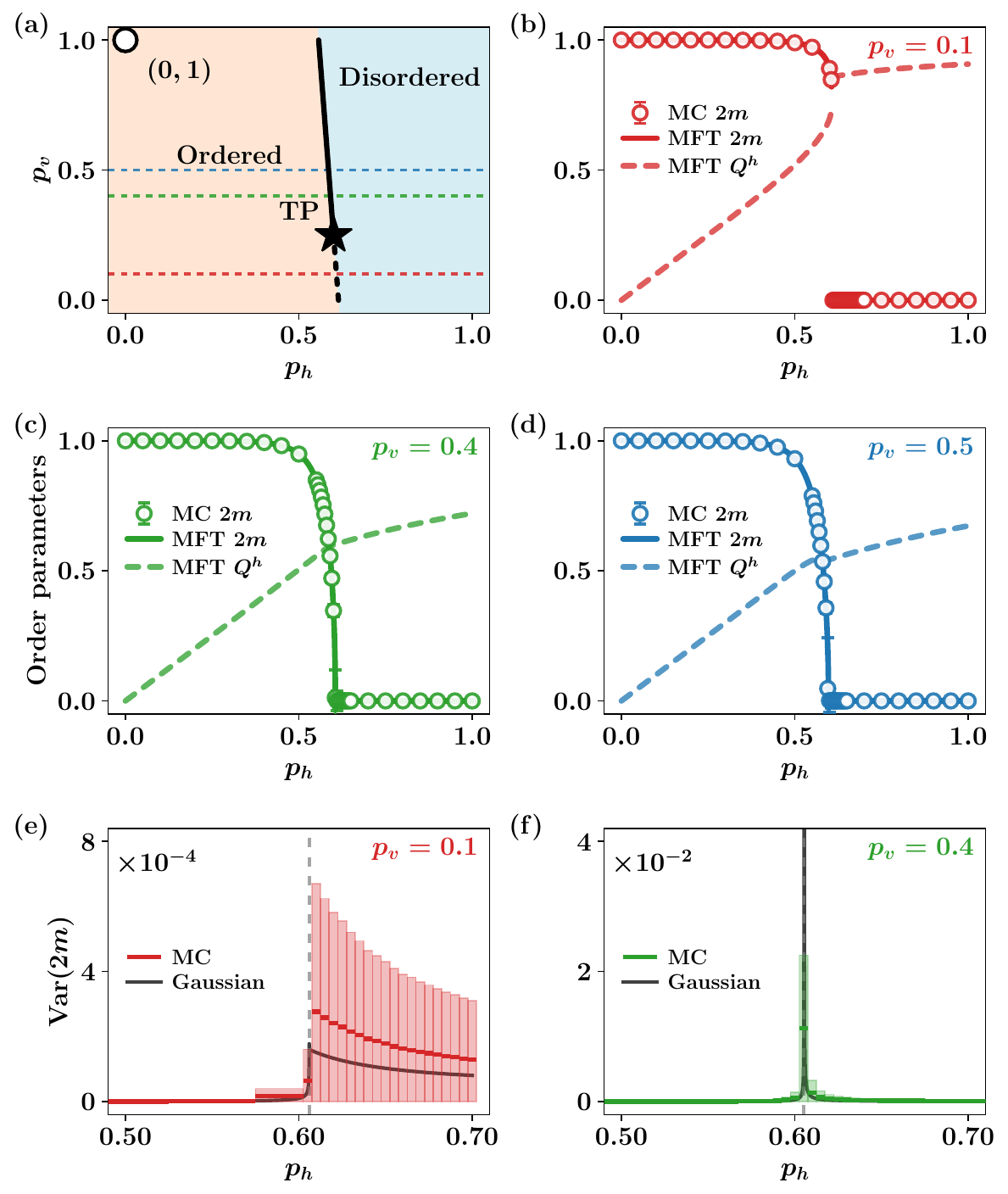}
\caption{(a) Phase diagram in the $(p_h,p_v)$ plane. Dashed and solid curves mark hybrid and continuous transitions, respectively. The star marks the tricritical point at $(p_h^t,p_v^t)\simeq(0.61,0.24)$; the open circle marks the baseline $(0,1)$. (b)\babelhyphen{--}(d) Solid lines, dashed lines, and circles show MFT $|m|$, MFT $Q^h$, and Monte Carlo (MC) $|m|$, respectively, at (b) $p_v=0.1$, (c) $0.4$, and (d) $0.5$. The transition is hybrid at $p_h^{**}\simeq 0.60$ for $p_v=0.1$ and continuous for $p_v=0.4$ and $0.5$. The hidden fraction is larger at $p_v=0.1$ over the entire range of $p_h$. (e) and (f) Order\hyph{}parameter variance $\mathrm{Var}(m)$: Gaussian fluctuation predictions (lines) and MC results (symbols). (e) At $p_v=0.1$, the predicted variance diverges as $(p_h^{**}-p_h)^{-1/2}$ on the ordered side and remains finite after the jump on the symmetric side. Coexistence causes deviations near the jump. (f) At $p_v=0.4$, the predicted variance diverges as $|p_h-p_h^{c}|^{-1}$ on both sides and agrees with the MC results away from the finite\hyph{}size peak. Vertical dashed lines in (e) and (f) mark the MFT transition points. All MC simulations used $N=16\times 10^4$.}
\label{fig:phase_diagram}
\end{figure}

Using mean\hyph{}field theory (MFT), we analytically show that the shy voter model exhibits phase transitions in the $(p_h,p_v)$ plane at finite transition points. In the large\hyph{}$p_v$ (low\hyph{}concealment) regime, a continuous transition line separates the ordered and symmetric states. In the small\hyph{}$p_v$ (high\hyph{}concealment) regime, by contrast, the transition is hybrid: the order parameter changes discontinuously across the transition, while the susceptibility diverges on the ordered side but remains finite on the symmetric side. The continuous and hybrid transition lines meet at a tricritical point. The hybrid transition is characterized by a set of critical exponents distinct from those of the continuous transition. The resulting phase diagram is shown in Fig.~\ref{fig:phase_diagram}(a). 





{\it Model:} Each node has an identical degree $k$, and the $k$ neighbors are not fixed; instead, they are randomly selected from the entire system at each update (annealed neighborhoods). We set $k=8$. This rule reflects a situation in which a voter may change their preferred candidate based on information received from $k$ individuals who change over time.

Each node $i$ can be in one of three possible states, $s_i \in \{+1,-1,0\}$, where $\pm1$ represent the two opposite visible states and $0$ represents the hidden state. Each update proceeds in three stages:
\begin{enumerate}
\item[(i)] Select a node $i$ uniformly at random. Then, choose $k$ neighbors uniformly at random from the remaining $N-1$ nodes. Among these $k$ neighbors, let $v$ denote the number of nodes in the visible states and $n_h=k-v$ the number of nodes in the hidden state. Among the $v$ visible neighbors, $v_+$ and $v_-$ are in the visible states $+1$ and $-1$, respectively. Thus, $v=v_++v_-$.

\item[(ii)] If $\max(v_+,v_-)>\ceil*{v/2}$, node $i$ is updated to the hidden state $s_i=0$ with probability $p_h$. With the remaining probability $1-p_h$, it is updated to the visible state corresponding to the majority among the $v$ visible neighbors.

\item[(iii)] Otherwise, node $i$ is updated to the hidden state $s_i=0$ with probability $1-p_v$, while each of the visible states $+1$ and $-1$ is chosen with probability $p_v/2$. We refer to this case hereafter as \emph{free choice}.

\end{enumerate}

We interpret each stage as follows: Stage (ii) applies when the neighborhood exhibits a dominant visible opinion. Here, $p_h$ quantifies the tendency of a voter to conceal their opinion when surrounded by a dominant visible opinion. 
Stage (iii) applies when the visible opinions in the neighborhood are nearly evenly split. Here, $p_v$ quantifies the tendency of a voter to express a visible opinion when the visible opinions are nearly evenly split.
The hidden state is not counted in the visible majority. This distinguishes the model from symmetric multistate voter models, in which all opinion labels play equivalent roles~\cite{herrerias2019multistate}. The constrained voter model also has nonequivalent states, but its centrist state participates in interactions with both extremist opinions~\cite{vazquez2003constrained,vazquez2004fate}. 

A small value of $p_v$ allows a node in the hidden state to remain hidden with high probability. Because hidden nodes contribute no visible opinion to their neighbors, they reduce the number of visible neighbors and thereby increase the likelihood of nearly evenly split neighborhoods. This concealment feedback sustains a large hidden fraction. A large value of $p_v$, by contrast, suppresses this feedback and reduces the hidden fraction. In our numerical analysis, we use $p_v=0.1$ and $0.4$ as representative values of the high\hyph{} and low\hyph{}concealment regimes, respectively; $p_v=0.5$ is included as an additional check of the low\hyph{}concealment behavior. When $(p_h,p_v)=(0,1)$, the model reduces to the conventional majority\hyph{}rule voter model.




\label{sec:mft}

{\it Mean\hyph{}field solution:}
Annealed sampling permits a large\hyph{}population deterministic description of the update process~\cite{vankampen2007stochastic,gardiner2009stochastic,kurtz1970ode}. Degree\hyph{}weighted neighbor probabilities are standard in network mean\hyph{}field treatments~\cite{dorogovtsev2008critical,sood2005heterogeneous,sood2008heterogeneous}.

Let $q_k^{\pm}$ and $q_k^h$ denote the probabilities that a node of degree $k$ occupies state $\pm1$ or $0$, respectively. 
Let $Q_i^{\alpha}$ denote the probability that a neighbor $j$ of node $i$ is in state $\alpha$, where $\alpha\in\{+,-,h\}$ corresponds to $s=+1,-1,0$. Since neighbor $j$ is in state $\alpha$ with probability $q_{k_j}^{\alpha}$,
\begin{align}
Q_i^{\alpha}=\sum_{j=1}^{N-1}\frac{k_j P_d(k_j)}{N\langle k \rangle}\,q_{k_j}^{\alpha}=q_k^{\alpha},
\label{eq:Q_neighbor}
\end{align}
where the last equality holds for large $N$ because $P_d(k^\prime)=\delta_{k^\prime,k}$ and $\langle k \rangle=k$.

Let $\psi_{k_j}^{\alpha}$ denote the probability that a node $j$ of degree $k_j$ updates to state $\alpha\in\{+,-,h\}$. Then, the rate equations are
\begin{align}
\dot q_{k_j}^{\alpha}&=-q_{k_j}^{\alpha}+\psi_{k_j}^{\alpha},
\label{eq:rate_eq_q}\\
\dot Q_i^{\alpha}&=-Q_i^{\alpha}+\Psi_i^{\alpha}(Q_i^+,Q_i^-,Q_i^h),
\label{eq:rate_eq_Q}
\end{align}
where
\begin{align}
\Psi_i^{\alpha}=\sum_{j=1}^{N-1}\frac{k_jP_d(k_j)}{N\langle k\rangle}\,\psi_{k_j}^{\alpha}.
\label{eq:Psi_neighbor}
\end{align}
In the steady state, $Q^{\alpha}=\Psi^{\alpha}$. Henceforth, because the system is homogeneous with respect to the node label $i$, we drop the node index. For a fixed degree, $\Psi^{\alpha}=\psi_k^{\alpha}$. The update rule allows $\Psi^{\alpha}$ to be written in polynomial form as a function of $(p_h,p_v,Q^+,Q^-,Q^h)$; the explicit derivation is given in the Supplemental Material (SM)~\cite{SM}.

Owing to the $\mathbb{Z}_2$ symmetry under the exchange $+1\leftrightarrow-1$, the system has a symmetric state given by $Q_0^+=Q_0^-=({1-Q_0^h})/2$, where the subscript $0$ denotes the symmetric fixed point. 
More generally, the state probabilities can be written as 
\begin{align}
Q^{\pm}=\frac{1-Q^h}{2}(1\pm2m),
\label{eq:Qpm_x}
\end{align}
where $m=(1/N_v)\sum_{i:\,s_i\neq0} s_i/2$, with $N_v$ denoting the number of nodes in the visible states $\pm1$. The quantity $m$ satisfies $-1/2 \le m \le 1/2$, measures the imbalance between the normalized populations of the $+1$ and $-1$ states, and serves as the order parameter for the visible states. In the steady state, the system satisfies the following coupled self\hyph{}consistency relations,
\begin{align}
F_1 & \equiv Q^h-\Psi^h(Q^h,m)=0, \label{eq:visible_equiv}\\
F_2 & \equiv 2(1-Q^h)m-w(Q^h,m)=0. 
\label{eq:visible_state}
\end{align}
Here, $w$ denotes the mean signed state $\langle s\rangle$ after the update. 
The explicit form of Eq.~\eqref{eq:visible_state} is provided in the SM. We solve Eq.~\eqref{eq:visible_equiv} for $Q^h$ as a function of $m$ and substitute the result into Eq.~\eqref{eq:visible_state} to obtain an equation for $m$. Defining $y\equiv Q^h-Q_0^h$, expanding $y$ as a power series in $m$, and substituting the resulting expansion into $F_2$, we obtain  
$G(m) \equiv \sum_{n=0}^7C_nm^n=0.$
The $\mathbb{Z}_2$ symmetry requires all even\hyph{}order coefficients to vanish. The odd\hyph{}order coefficients depend on $p_h$ and $p_v$ both explicitly and through $Q_0^h$.

In equilibrium statistical physics, macroscopic single\hyph{}variable free\hyph{}energy landscapes provide a useful framework for understanding phase transitions and critical phenomena~\cite{jang2015ashkin,kim2024entropy,kim2026heterogeneous}. This landscape approach has also been extended to nonequilibrium and nonlinear systems~\cite{jhun2022quantum,song2020effective}. Here, we define an {\it ad hoc} effective potential by integrating $G(m)$. With $G(m)$ truncated at fifth order, the potential takes the sixth\hyph{}order Ginzburg\babelhyphen{--}Landau form used in tricritical theory~\cite{lawrie1984tricritical,goldenfeld2018lectures},
\begin{align}
\mathcal{L}(m)=\int G(m)\,dm \equiv \tfrac12C_1m^2+\tfrac14C_3m^4+\tfrac16C_5m^6.
\label{eq:landscape}
\end{align}
Stationary states satisfy $d\mathcal{L}/dm=0$, and stable states correspond to local minima of $\mathcal{L}(m)$.

\begin{figure}[t]
\centering
\includegraphics[width=\columnwidth]{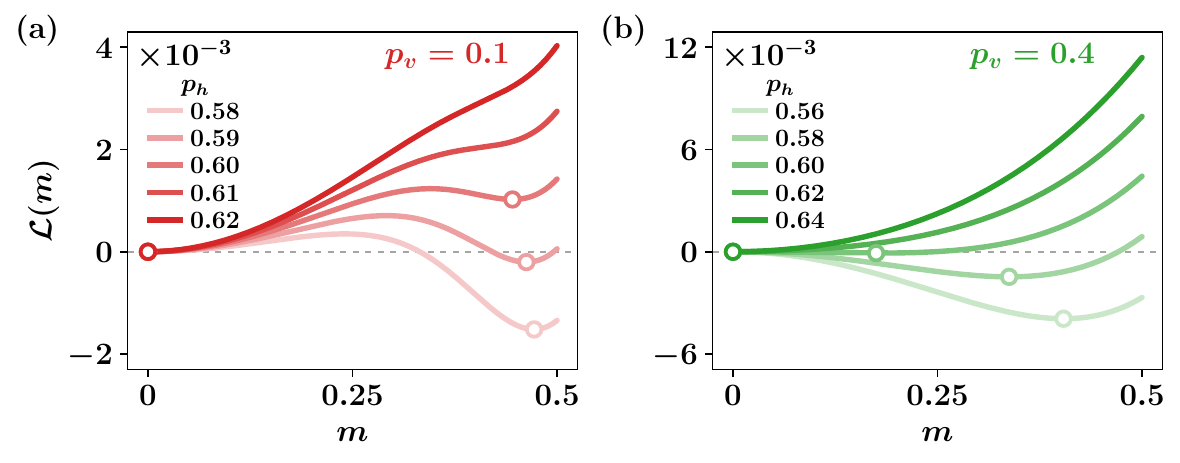}
\caption{Landscape of the {\it ad hoc} Ginzburg\babelhyphen{--}Landau potential $\mathcal{L}(m)$ for $k=8$. (a) At $p_v=0.1$, the minima of the ordered state coexist with the symmetric minimum at $m=0$. The global minimum switches to $m=0$ at $p_h^*\simeq 0.59$, and the ordered local minima disappear at the spinodal point $p_h^{**}\simeq 0.60$, where the hybrid transition occurs. (b) At $p_v=0.4$, the global minima (marked by circles) approach $m=0$ continuously as $p_h$ increases, generating a continuous transition.}
\label{fig:GL}
\end{figure}

%

The behavior of $\mathcal{L}(m)$ is shown in Fig.~\ref{fig:GL}. 
When $p_v=0.1$ [Fig.~\ref{fig:GL}(a)], global minima (marked by circles) occur at finite $m_c$, at which $\mathcal{L} < 0$, $d\mathcal{L}/dm=0$, and $d^2\mathcal{L}/dm^2 > 0$. The positive value $m_c$ decreases with increasing $p_h$. However, the ordered minima cease to be global when $p_h > p_h^* \simeq 0.59$. The global minimum thus switches from $m=\pm m_c$ to $m=0$ at $p_h^*$. In this coexistence regime, $C_1 >0$, $C_3 <0$, and $C_5 >0$. 

For $p_h^*<p_h<p_h^{**}$, the ordered minima persist as metastable local minima with $d^2\mathcal{L}/dm^2 > 0$. The point $p_h^{**}$ is where $d\mathcal{L}/dm=0$ and $d^2\mathcal{L}/dm^2=0$. This property characterizes a mean\hyph{}field spinodal point~\cite{binder1987firstorder}, which we identify from the condition $\det[\partial(F_1,F_2)/\partial(Q^h,m)]=0$.    
We examine the susceptibility at $p_h^{**}$ and at the continuous transition point $p_h^{c}$ below. 

When $p_v=0.4$ [Fig.~\ref{fig:GL}(b)], the order parameter is determined by the global minimum $m_c$ of $\mathcal{L}(m)$, and $m_c$ decreases continuously with increasing $p_h$. When $p_h$ reaches $p_h^{c}$, $m_c$ becomes zero. In the ordered state, $C_1< 0$, $C_3 > 0$, and $C_5>0$. Thus, a continuous transition occurs. In general, the transition at $C_1=0$ is continuous for $C_3>0$ and hybrid for $C_3<0$.  


The small\hyph{}$m$ expansion determines the mean\hyph{}field critical exponent of the order parameter. For the continuous transition, the leading terms in $\mathcal{L}(m)$ are $C_1m^2$ and $C_3m^4$, while the higher\hyph{}order term $C_5m^6$ can be neglected. Balancing these two leading terms gives $m\sim |C_1|^{1/2}$, yielding the mean\hyph{}field order\hyph{}parameter exponent $\beta=1/2$. At the tricritical point, where $C_3=0$, balancing $C_1m^2$ and $C_5m^6$ gives $m\sim|C_1|^{1/4}$, yielding $\beta=1/4$.



The tricritical point, where the continuous and hybrid transition lines meet, is determined by the criterion $C_3=0$ on the stability line. It is numerically estimated to be at $p_h^t \simeq 0.61$, $p_v^t \simeq 0.24$, and $Q_0^{h,t} \simeq 0.71$. Thus, for $p_v=0.1 < p_v^t$, corresponding to the high\hyph{}concealment regime, the transition is hybrid, whereas for $p_v=0.4 > p_v^t$, corresponding to the low\hyph{}concealment regime, the transition is continuous.


{\it Numerical simulations:} Monte Carlo (MC) simulations are performed with $k=8$. 
At each update, $k$ neighbors are randomly chosen using annealed neighborhoods. We measure the order parameter $m$ and the hidden fraction $Q^h$. Both the MC simulations and the MFT start from the initial condition $(m,Q^h)=(1/2,0)$, and $p_h$ is then increased. The system size is $N=16\times10^4$. 
Details of the simulations are provided in the SM.


Figure~\ref{fig:phase_diagram}(b)\babelhyphen{--}(d) shows the behavior of the order parameter $m$ and the hidden fraction $Q^h$. At $p_v=0.1$, both $m$ and $Q^h$ jump, whereas at $p_v=0.4$ and $0.5$, both quantities vary continuously across the transition. The MFT and MC results agree on the transition point, transition type, and steady\hyph{}state values, supporting the validity of the MFT.

Fluctuations around a stable steady state are described by a Gaussian linear\hyph{}noise approximation~\cite{vankampen2007stochastic,gardiner2009stochastic}, as derived in the SM. The order\hyph{}parameter variance scales as $\mathrm{Var}(m)\propto {1}/{\det A}$,
where $A=I-J_T$, $I$ is the identity matrix, and $J_T$ is the one\hyph{}step Jacobian (see the SM for its explicit form). The susceptibility $\chi$ to an external field conjugate to $m$ contains the same singular factor, $\chi\propto 1/\det A$ (see the SM). Thus, $\mathrm{Var}(m)$ and $\chi$ diverge with the same critical exponent, the susceptibility exponent $\gamma$, defined by $\chi\propto|p_h-p_h^{\mathrm{tr}}|^{-\gamma}$, where $p_h^{\mathrm{tr}}$ denotes the transition point. The value of $\gamma$ is therefore determined by how $\det A$ approaches zero at $p_h^{\mathrm{tr}}$.

For $p_v=0.1$, the ordered state persists as a metastable state up to the spinodal point $p_h^{**}$, where it loses stability and the system undergoes a hybrid transition. As the spinodal point is approached along the ordered state, the order\hyph{}parameter variance diverges as $\mathrm{Var}(m)\propto (p_h^{**}-p_h)^{-1/2}$, yielding $\gamma=1/2$ (see the SM for details). By contrast, the symmetric state with $m=0$ reached after the jump is stable and has a finite variance. This asymmetric behavior of the order\hyph{}parameter variance is a feature of the hybrid transition. A discontinuous jump accompanied by critical singularities is also found in $k$\hyph{}core percolation~\cite{dorogovtsev2008critical,dorogovtsev2006kcore,goltsev2006kcore}; in the present model, the singularity arises at the loss of metastability of the ordered state.


For $p_v=0.4$, the order parameter decreases continuously to zero at the continuous transition point $p_h^{c}$, where the ordered and symmetric states meet. The variance diverges as $\mathrm{Var}(m)\propto|p_h-p_h^{c}|^{-1}$ on both sides of the transition, yielding $\gamma=1$. The variance reaches order $10^{-2}$ only near the continuous transition point. The MFT and MC results agree across the transition, apart from the finite\hyph{}size peak near the continuous transition point [Fig.~\ref{fig:phase_diagram}(e) and (f)].

{\it Nonequilibrium behavior:} In the shy voter model, the update rule can break detailed balance, producing nonzero probability currents even in the steady state. The relation between probability currents, irreversibility, and entropy production is reviewed in Refs.~\cite{seifert2012stochastic,schnakenberg1976network}. Entropy production has also been investigated in majority\hyph{}vote and cooperative opinion models~\cite{crochik2005entropy,tome2012entropy,hawthorne2023thermodynamics,oliveira2024entropy}. We define the entropy\hyph{}flux observable considered here as

\begin{widetext}
\begin{align}
\Sent &=
\sum_{\mathbf n}P(\mathbf n)
\sum_{s(t)}P\bigl(s(t)\bigr)
\sum_{s(t+1)\neq s(t)}
P\bigl(s(t+1)\mid\mathbf n\bigr)
\ln
\frac{
P\bigl(s(t+1)\mid\mathbf n\bigr)
}{P\bigl(s(t)\bigr)},
\label{eq:entropy_flux}
\end{align}
\end{widetext}
where $\mathbf n=(v_+,v_-,n_h)$ gives the numbers of neighbors in states $+1$, $-1$, and $0$, respectively. The probability of sampling this neighborhood composition, $P(\mathbf n)$, follows a multinomial distribution. In the steady state, $P\bigl(s(t)\bigr)$ is the probability that the selected node occupies state $s(t)$ before the update. The conditional probability $P\bigl(s(t+1)\mid\mathbf n\bigr)$ gives the probability that the node adopts state $s(t+1)$ after the update, given the sampled neighborhood. The sum includes only updates that change the node's state, $s(t+1)\neq s(t)$. This observable differs from the standard entropy flux, whose logarithm contains the ratio of forward to reverse transition rates~\cite{seifert2012stochastic,schnakenberg1976network}. Here, the logarithm compares the post\hyph{}update probability with the steady\hyph{}state probability of the pre\hyph{}update state. We therefore use $\Sent$ as a measure of state\hyph{}changing activity and do not identify it with the thermodynamic entropy production. A derivation is provided in the SM.

\begin{figure}[t]
\includegraphics[width=\columnwidth]{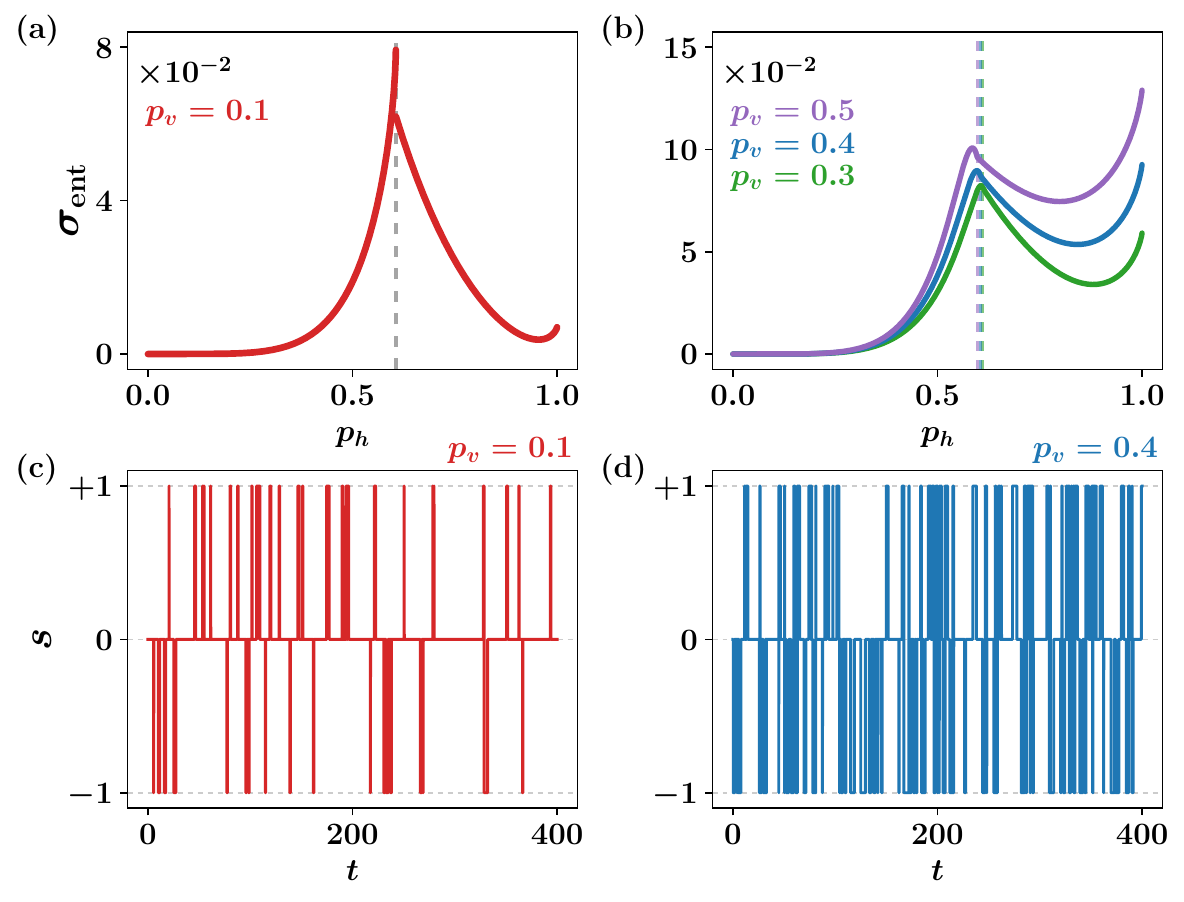}
\caption{Entropy flux $\Sent$ versus $p_h$ for (a) $p_v=0.1$ (high\hyph{}concealment regime) and (b) $p_v=0.3$, $0.4$, and $0.5$ (low\hyph{}concealment regime).
Vertical dashed lines mark the spinodal point $p_h^{**}$ in (a) and the continuous transition points $p_h^{c}$ in (b); each line in (b) has the color of the corresponding $p_v$.
To the right of each dashed line, the system is in the symmetric state with $m=0$.
State\hyph{}changing updates between the visible and hidden states nevertheless continue, so $\Sent$ remains finite.
In (a), $\Sent$ jumps at $p_h^{**}$ and then decreases to a small value;
in (b), $\Sent$ remains large in the symmetric state and increases with $p_v$. (c) and (d) State evolution $s_i(t)$ of a representative node $i$ in MC simulations for (c) $(p_h,p_v)=(0.7,0.1)$ and (d) $(0.7,0.4)$. The time series $s_i(t)$ are shown over the last $10^3$ MC sweeps after reaching the steady state. For $p_v=0.1$, the node switches only infrequently from the hidden state $s=0$. In contrast, for $p_v=0.4$, the node switches frequently between the hidden and visible states. Here, $N=16\times10^4$. 
}
\label{fig:entropy}
\end{figure}

In the ordered state, the entropy fluxes are comparable in the high\hyph{} and low\hyph{}concealment regimes. However, they differ markedly after the system enters the symmetric state with $m=0$, i.e., to the right of the vertical dashed lines in Fig.~\ref{fig:entropy}(a) and (b). For $p_v=0.1$ (high\hyph{}concealment regime), the system jumps to the symmetric state at the spinodal point $p_h^{**}$. For $p_v=0.4$ (low\hyph{}concealment regime), it reaches the symmetric state continuously at $p_h^{c}$. In the symmetric state, even though $m$ remains zero, the entropy flux is governed primarily by the steady\hyph{}state hidden fraction and the rate of state\hyph{}changing updates.

At $p_h=1$, the relation between concealment and entropy flux becomes particularly clear. Under the update rule, stage (ii) always assigns the hidden state, whereas stage (iii) assigns a visible state with probability $p_v$. The steady\hyph{}state visible fraction therefore satisfies
$1-Q^h=p_v\mathcal{P}_{\mathrm{free}}$, where $\mathcal{P}_{\mathrm{free}}$ is the steady\hyph{}state probability that the sampled neighborhood satisfies the free\hyph{}choice condition of stage (iii). 
State\hyph{}changing updates between the visible and hidden states therefore depend on the steady\hyph{}state visible fraction. 

When $p_v=0.1$ (high\hyph{}concealment regime) in the symmetric state, the small value of $p_v$ yields a small visible fraction and, consequently, a small entropy flux.
When $p_v=0.4$ (low\hyph{}concealment regime) in the symmetric state, the larger value of $p_v$ yields a larger visible fraction and, consequently, a larger entropy flux.

Although both symmetric states have $m\simeq 0$, they differ in the rate of state\hyph{}changing updates. Thus, the entropy flux characterizes nonequilibrium dynamics in the shy voter model beyond what is captured by the order parameter $m$.

{\it Summary and discussion:} 
The shy voter model extends majority\hyph{}rule voter dynamics by introducing a hidden state that is fundamentally distinct from the visible states. Unlike the states in multistate generalizations, which play equivalent roles, the hidden state does not contribute to the majority count, yet it shapes the nonequilibrium dynamics, as reflected in the entropy flux. This asymmetry, arising from the psychological tendency to conceal opinions, generates a rich phase diagram over a finite range of control parameters. In particular, the visibility probability \(p_v\) regulates the steady\hyph{}state hidden fraction and thereby determines the transition type, the order\hyph{}parameter variance, the susceptibility, and the entropy flux.

The high\hyph{} and low\hyph{}concealment regimes exhibit distinct singular behavior. In the former regime, the ordered state persists as a metastable state up to the spinodal point $p_h^{**}$, where it loses stability and the system jumps discontinuously to the symmetric state. As the spinodal point is approached along the ordered state, the susceptibility diverges as $\chi\propto (p_h^{**}-p_h)^{-1/2}$. This divergence occurs only on the ordered side and ends at the jump. In the latter regime, the order parameter instead decreases continuously to zero at the continuous transition point $p_h^{c}$, with $\chi\propto|p_h-p_h^c|^{-1}$ on both sides of the transition. Thus, the susceptibility exponents are $\gamma=1/2$ at the spinodal point and $\gamma=1$ at the continuous transition point. The order\hyph{}parameter exponent is $\beta=1/2$ along the continuous transition line and $\beta=1/4$ at the tricritical point.

The entropy flux provides an additional probe of the system's nonequilibrium dynamics. In the ordered state, the high\hyph{} and low\hyph{}concealment regimes exhibit comparable entropy fluxes. After the transition to the symmetric state, however, the high\hyph{}concealment regime has a smaller visible fraction and a lower rate of state\hyph{}changing updates, resulting in a smaller entropy flux than in the low\hyph{}concealment regime. Thus, states with the same order parameter, $m=0$, can nevertheless differ in their hidden fractions and nonequilibrium dynamics. At $(p_h,p_v)=(0,1)$, concealment is absent, and the majority\hyph{}rule voter model evolves toward an absorbing consensus without undergoing a phase transition. Introducing concealment therefore qualitatively changes the transition structure, critical behavior, and nonequilibrium dynamics.


Continuous and discontinuous transition lines also meet in three\hyph{}state equilibrium models~\cite{lawrie1984tricritical,blume1966ising,capel1966anisotropy,kim2024entropy}. Changes in transition order occur in nonlinear voter~\cite{castellano2009nonlinear,nyczka2012qvoter,peralta2018nonlinear} and inertial majority\hyph{}vote models~\cite{chen2017inertia}. Our results show that opinion concealment provides another mechanism for changing the transition order.

The present study considers the simplest setting: two visible states and one hidden state, with annealed neighborhoods of fixed degree $k$ in the mean\hyph{}field limit. Real\hyph{}world social systems~\cite{castellano2009social,redner2019review,dorogovtsev2008critical} have three features that this setting omits: voters choose among more than two visible opinions, differ in the number of neighbors, and interact through multiple social layers. In spin models, each of these features has been shown to alter the universality class: more than two spin states~\cite{potts1952some,wu1982potts}, heterogeneous numbers of neighbors~\cite{jang2015ashkin,kim2025ashkin,kim2026heterogeneous}, and multiple coupled layers~\cite{jang2015ashkin,kim2025ashkin,kim2021link}. Including these features in the shy voter model would reveal whether the features also alter the phase diagram and universality class of the model.

In this study, we show that opinion concealment can substantially alter collective dynamics and phase behavior in the shy voter model. Concealment changes the transition order, modifies critical behavior, and supports coexistence between ordered and symmetric states. Symmetric states with the same visible order parameter can differ in their hidden fractions and dynamical activity. These results highlight the importance of both visible opinions and hidden states for understanding collective opinion dynamics. These findings may also inspire future models that connect what voters think privately, what they express publicly, and what opinion polls report.



\section*{Data availability}
The code that supports the mean\hyph{}field results of this article is openly available~\cite{shyvoter_code}. The Monte Carlo data are available upon reasonable request from the authors.

\begin{acknowledgments}
The authors thank Prof. B. Min for helpful discussions. 
This work was supported by the National Research Foundation of Korea (Grant No. RS-2023-00279802 and No. RS-2026-25556154). 
\end{acknowledgments}

%
\end{bibunit}

\clearpage
\newpage

\onecolumngrid

\begin{bibunit}[apsrev4-2]
\SetBibliographyUnit{.sm}{S}

\setcounter{section}{0}
\setcounter{equation}{0}
\setcounter{figure}{0}
\setcounter{table}{0}
\renewcommand{\thesection}{S\arabic{section}}
\renewcommand{\theequation}{S\arabic{equation}}
\renewcommand{\thefigure}{S\arabic{figure}}
\renewcommand{\theHfigure}{S\arabic{figure}}
\renewcommand{\thetable}{S\arabic{table}}

\begin{center}
{\large\bfseries Supplemental Material}\\[4pt]
{\itshape A Shy Voter Model: Entropy Flux in Opinion Formation}\\[2pt]
Cook Hyun Kim, Sangjoon Park, and B. Kahng
\end{center}

\vspace{6pt}

\section{Mean-field equations and expansion coefficients}
\label{app:coefficients}

This Supplemental Material details the mean\hyph{}field construction for $k=8$, obtained by averaging the microscopic update rule over the annealed distribution of neighborhood compositions. The deterministic limit and fluctuation expansion follow the general framework for population jump processes~\cite{vankampen2007stochastic,gardiner2009stochastic,kurtz1970ode}.

\subsection{Hidden\hyph{}state equation}

For a node $i$ of degree $k$, let $n_h$ and $v$ denote the numbers of hidden and visible neighbors, respectively, so that $v=k-n_h$. Among the $v$ visible neighbors, $v_+$ are in state $+1$, and the remaining $v_-=v-v_+$ are in state $-1$.

Let $q_k^{\pm}$ and $q_k^h$ denote the probabilities that a node of degree $k$ occupies state $\pm1$ or $0$, respectively.
Let $Q_i^{\alpha}$ denote the probability that a neighbor of node $i$ is in state $\alpha\in\{+,-,h\}$. A neighbor $j$ is in state $\alpha$ with probability $q_{k_j}^{\alpha}$. Then, $Q_i^{\alpha}=\sum_{j=1}^{N-1}\frac{k_jP_d(k_j)}{N\langle k \rangle}q_{k_j}^{\alpha}$. We write $Q_i^{\pm}=(1-Q_i^h)(1\pm2m_i)/2$. Here, $m_i=0$ when the populations in states $+1$ and $-1$ are equal; otherwise, $m_i\neq0$. The following relation holds:
\begin{align}
\frac{Q_i^{\pm}}{Q_i^++Q_i^-} = \frac{(1-Q_i^h)(1\pm2m_i)/2}{1-Q_i^h}
= \frac12 \pm m_i.
\nonumber
\end{align}
In what follows, we omit the node subscript $i$, since all nodes have the same degree and their neighbors are selected uniformly at random. The hidden fraction cancels from the conditional probability, so the order parameter $m$ alone sets the imbalance among visible neighbors. The $v$ visible neighbors are selected independently, so the composition $(v_+,v_-)$ of the states $+1$ and $-1$ occurs with probability
$$\binom{v}{v_+}\left(\tfrac12+m\right)^{v_+}\left(\tfrac12-m\right)^{v_-}.$$

Each neighborhood composition is assigned the concealment probability specified by the corresponding update rule.
\begin{align}
\omega_h(v_+,v) = \begin{cases}
p_h, & v_+ > \ceil*{v/2} \quad (+1 \text{ dominant}), \\[2pt]
p_h, & v_- > \ceil*{v/2} \quad (-1 \text{ dominant}), \\[2pt]
1-p_v, & \text{otherwise} \quad (+1 \text{ and } -1 \text{ states nearly evenly split}).
\end{cases}
\nonumber
\end{align}
The $\mathbb{Z}_2$ symmetry assigns the same weight $p_h$ to the two dominant cases. The two cases therefore combine into the single condition of stage (ii), $\max(v_+, v_-)>\ceil*{v/2}$. Let $\Psi^h$ be the probability that a node updates to the hidden state, averaged over the number $n_h$ of hidden neighbors:
\begin{align}
    \Psi^h = \sum_{n_h=0}^{k}\binom{k}{n_h}(Q^h)^{n_h}
             (1-Q^h)^{k-n_h}\,\Sigma^h(k-n_h),
    \label{eq:psi_h_full}
\end{align}
where $\Sigma^h(v)$, with $v=k-n_h$, is the concealment probability $\omega_h$ averaged over the compositions of the $v$ visible neighbors. This probability is given as
\begin{align}
\Sigma^h(v) &= \sum_{v_+=0}^{v} \binom{v}{v_+}
\left(\tfrac12+m\right)^{v_+}
\left(\tfrac12-m\right)^{v-v_+} \omega_h(v_+,v).\label{eq:sigma_h}
\end{align}

For $k=8$, Eq.~\eqref{eq:psi_h_full} is a degree\hyph{}eight polynomial in $Q^h$. The coefficients depend on $p_h$, $p_v$, and $m^2$. The $\mathbb{Z}_2$ symmetry excludes odd\hyph{}power terms in $m$. The simulation code evaluates Eqs.~\eqref{eq:psi_h_full}\babelhyphen{--}\eqref{eq:sigma_h}. The fully expanded polynomial contains no additional information and is omitted. Solving $Q^h=\Psi^h$ at $m=0$ yields the hidden fraction $Q_0^h(p_h,p_v)$ of the main text, and the dependence of $Q_0^h$ on $p_v$ arises from the concealment feedback described in the model section of the main text.

\subsection{Expansion coefficients}

A series expansion of the steady\hyph{}state conditions at small $m$ fixes the transition type. Setting $Q^h=Q_0^h+y$ in $Q^h-\Psi^h=0$ leaves an equation determining $y$ as a function of $m$. An asymptotic solution about $m=0$ to fifth order gives
\begin{align}
    y(m) = G_2 m^2 + G_4 m^4 + \mathcal{O}(m^6).
\end{align}
The $\mathbb{Z}_2$ symmetry removes the odd\hyph{}power coefficients.

The neighborhood average also determines the visible\hyph{}state equation. Writing $\mathcal{M}(v)=\{v_+ : \max(v_+,v-v_+)>\ceil*{v/2}\}$ for the set of compositions in stage (ii),
\begin{align}
    \Sigma^{\pm}(v) = (1-p_h) \sum_{v_+ \in \mathcal{M}(v)}
        \binom{v}{v_+}
        \left(\tfrac12+m\right)^{v_+}
        \left(\tfrac12-m\right)^{v-v_+}
        \mathrm{sgn}(2v_+-v),
    \label{eq:sigma_pm}
\end{align}
and the mean signed state after the update is
\begin{align}
    w(Q^h,m) = \sum_{n_h=0}^{k}\binom{k}{n_h}(Q^h)^{n_h}(1-Q^h)^{k-n_h}
           \Sigma^{\pm}(k-n_h).
    \label{eq:sprime_full}
\end{align}
Stage (iii) contributes zero to $w$ at zero field because free choice selects $+1$ and $-1$ with equal probability. Substituting $y(m)$ into $F_2=0$ of Eq.~\eqref{eq:visible_state} and expanding in $m$ yields the coefficients $C_n$ quoted in the main text, with $C_0=C_2=C_4=0$ identically.

\subsubsection{Explicit coefficients at $k=8$}

The surviving coefficients are rational functions of $(p_h,p_v,Q_0^h)$. Three recurring polynomials are
\begin{align}
    A_1 &= 19 - 35Q_0^h + 315(Q_0^h)^2 - 1155(Q_0^h)^3 + 2625(Q_0^h)^4 - 3633(Q_0^h)^5 + 2849(Q_0^h)^6 - 1001(Q_0^h)^7,
    \label{eq:A1} \\
    A_2 &= -5 + 5Q_0^h - 45(Q_0^h)^2 + 165(Q_0^h)^3 - 375(Q_0^h)^4 + 519(Q_0^h)^5 - 407(Q_0^h)^6 + 143(Q_0^h)^7,
    \label{eq:A2} \\
    A_3 &= -5 + 65Q_0^h - 405(Q_0^h)^2 + 1385(Q_0^h)^3 - 2855(Q_0^h)^4 + 3531(Q_0^h)^5 - 2431(Q_0^h)^6 + 715(Q_0^h)^7 .
    \label{eq:A3}
\end{align}
The common denominator of the expansion is
\begin{align}
    D = 19 - 455Q_0^h + 2835(Q_0^h)^2 - 9695(Q_0^h)^3 + 19985(Q_0^h)^4 - 24717(Q_0^h)^5 + 17017(Q_0^h)^6 - 5005(Q_0^h)^7 + 7(p_h+p_v)A_3.
    \label{eq:Ddef}
\end{align}

The leading coefficient of $y(m)$ is
\begin{align}
    G_2 & \equiv \frac{14\,(p_h+p_v-1)\,(Q_0^h-1)^2\,B_2}{D},
    \cr
    B_2 & \equiv -5 - 30Q_0^h + 105(Q_0^h)^2 - 340(Q_0^h)^3 + 525(Q_0^h)^4 - 462(Q_0^h)^5 + 143(Q_0^h)^6.
    \label{eq:G2}
\end{align}
The coefficient $G_2$ is recorded in full. The next coefficient has the form
\begin{align}
    G_4 = \frac{14\,(p_h+p_v-1)\,(Q_0^h-1)^3\,B_4(Q_0^h;p_h,p_v)}{D^3},
    \label{eq:G4}
\end{align}
with $B_4$ of degree $19$ in $Q_0^h$, and quadratic in each of $p_h$ and $p_v$.

The coefficient that fixes the stability of the symmetric state is
\begin{align}
    C_1 = \frac{(Q_0^h-1)\left(A_1 + 7 p_h A_2\right)}{8}.
    \label{eq:C1_explicit}
\end{align}
The hidden fraction satisfies $Q_0^h<1$ wherever the order parameter is defined. The stability line $C_1=0$ therefore reduces to
\begin{align}
    A_1(Q_0^h) + 7 p_h A_2(Q_0^h) = 0.
    \label{eq:stability_line}
\end{align}
Equation~\eqref{eq:stability_line} is solved together with $Q_0^h=\Psi^h(Q_0^h;p_h,p_v)$. For $C_3>0$, the root of Eq.~\eqref{eq:stability_line} is the continuous transition line $p_h^{c}(p_v)$. For $C_3<0$, $p_h$ crosses the root while the ordered state remains stable. The hybrid transition then occurs at the spinodal point $p_h^{**}(p_v)$, determined by $\det[\partial(F_1,F_2)/\partial(Q^h,m)]=0$.

The coefficient that fixes the transition type is
\begin{align}
    C_3 = -\frac{7\,(Q_0^h-1)^2\,B_3(Q_0^h;p_h,p_v)}{2\,D},
    \label{eq:C3_explicit}
\end{align}
where $B_3$ has degree $13$ in $Q_0^h$, is quadratic in $p_h$, and is linear in $p_v$. The quadratic dependence enters through $p_h(p_h+p_v)$. Numerical evaluation gives $D<0$ at every transition point in the physical range $0<Q_0^h<1$, so $C_3$ and $B_3$ have the same sign. The sign of $C_3$ at the root of Eq.~\eqref{eq:stability_line} determines the transition type. Simultaneous zeros of $C_1$ and $C_3$ locate the tricritical point $(p_h^t,p_v^t)$.

The coefficient at fifth order is
\begin{align}
    C_5 = \frac{7\,(Q_0^h-1)^3\,B_5(Q_0^h;p_h,p_v)}{2\,D^3},
    \label{eq:C5_explicit}
\end{align}
with $B_5$ of degree $26$ in $Q_0^h$, quartic in $p_h$, and cubic in $p_v$. The coefficient $C_5$ becomes the leading nonlinearity at $C_3=0$. The coefficient $C_5$ therefore fixes the tricritical exponent $\beta=1/4$ and the local stability of the expansion.

The numerators $B_3$, $B_4$, and $B_5$ run to several lines and are not reproduced. The construction starts from Eqs.~\eqref{eq:psi_h_full}\babelhyphen{--}\eqref{eq:sigma_h}. Expanding $Q^h-\Psi^h=0$ about $Q_0^h$ gives $y(m)$. Substitution into $F_2=0$ of Eq.~\eqref{eq:visible_state} gives the coefficients of $m^n$. The computer\hyph{}algebra script for the expansion, including the full expressions for $B_3$, $B_4$, and $B_5$, and the numerical solver for the transition lines and the tricritical point are openly available~\cite{shyvoter_code}. Numerical evaluation of the unexpanded equations reproduces the reported results, including the tricritical point.

\section{Gaussian fluctuations from the Lyapunov equation}
\label{app:lyapunov}

\subsection{The one\hyph{}step map}

At the mean\hyph{}field level, let $(Q^h(t),m(t))$ denote the state entering the one\hyph{}step map and $(Q^h(t+1),m(t+1))$ its image. One update assigns a single node to one of the three states $s(t+1)\in\{+1,-1,0\}$ according to the neighborhood composition at step $t$. Two quantities specify the distribution of $s(t+1)$,
\begin{align}
    Q^h(t+1) = P\bigl(s(t+1)=0\bigr),
    \qquad
    w(t+1)\equiv\langle s(t+1)\rangle = P\bigl(s(t+1)=+1\bigr)-P\bigl(s(t+1)=-1\bigr).
\end{align}
Here, $w$ is the mean signed state over all nodes, including hidden nodes.

The two quantities are computed in two stages. First, fix the number $n_h$ of hidden neighbors, so that $v=k-n_h$ neighbors are visible. The first stage carries the binomial weight $\binom{k}{n_h}[Q^h(t)]^{n_h}[1-Q^h(t)]^{k-n_h}$. Second, sum over $v_+$, the number of visible neighbors in state $+1$, so that $v_-=v-v_+$ neighbors are in state $-1$. The second stage is the sum defined by $\Sigma^{h}(v)$ and $\Sigma^{\pm}(v)$ in Eqs.~\eqref{eq:sigma_h} and \eqref{eq:sigma_pm}. Combining the two stages,
\begin{align}
    Q^h(t+1)
    &= \sum_{n_h=0}^{k}\binom{k}{n_h}
    [Q^h(t)]^{n_h}[1-Q^h(t)]^{k-n_h}
    \Sigma^{h}(k-n_h),
    \label{eq:map_T}\\[2pt]
    w(t+1)
    &= \sum_{n_h=0}^{k}\binom{k}{n_h}
    [Q^h(t)]^{n_h}[1-Q^h(t)]^{k-n_h}
    \Sigma^{\pm}(k-n_h).
    \nonumber
\end{align}
The first line is Eq.~\eqref{eq:psi_h_full} written as the one\hyph{}step map, $Q^h(t+1)=\Psi^h[Q^h(t),m(t)]$. The second line is Eq.~\eqref{eq:sprime_full} written as the one\hyph{}step map, $w(t+1)=w[Q^h(t),m(t)]$. The order parameter $m$ is normalized by the visible fraction. Since $Q^{\pm}=(1-Q^h)(1\pm2m)/2$ [Eq.~\eqref{eq:Qpm_x}] gives $w=Q^+-Q^-=2(1-Q^h)m$,
\begin{align}
    m(t+1)
    =\frac{w(t+1)}{2\bigl[1-Q^h(t+1)\bigr]}.
    \label{eq:Phi_prime}
\end{align}

We therefore define the one\hyph{}step map as
\begin{align}
    T:\bigl(Q^h(t),m(t)\bigr)
    \mapsto
    \bigl(Q^h(t+1),m(t+1)\bigr).
\end{align}
At a fixed point,
\begin{align}
    Q^h(t+1)=Q^h(t)\equiv Q^h,
    \qquad
    m(t+1)=m(t)\equiv m,
\end{align}
and the fixed points of $T$ coincide with the roots of Eqs.~\eqref{eq:visible_equiv} and \eqref{eq:visible_state} and the mean\hyph{}field solutions of the main text.

\subsection{Jacobian and noise}
\label{app:noise}

\subsubsection{Linearized dynamics and Jacobian}

A fluctuation of $(Q^h,m)$ has two contributions: deterministic propagation and stochastic injection. Let $\xi_b$ denote the deviation from a fixed point at batch $b$, $f=1/\sqrt{N}$ the fraction of nodes updated per batch, and $\eta_b$ the stochastic increment accumulated during the batch.
\begin{align}
    \xi_{b+1}
    =\bigl(I-fA\bigr)\xi_b+\eta_b,
    \qquad
    A\equiv I-J_T,
    \label{eq:OU}
\end{align}
where $J_T$ is the one\hyph{}step Jacobian of the map $T$ evaluated at the fixed point.

For the one\hyph{}step map
\begin{align}
    \bigl(Q^h(t),m(t)\bigr)
    \mapsto
    \bigl(Q^h(t+1),w(t+1)\bigr),
\end{align}
we define
\begin{align}
    \Psi_Q
    &=\frac{\partial Q^h(t+1)}{\partial Q^h(t)},
    &\,
    \Psi_m
    &=\frac{\partial Q^h(t+1)}{\partial m(t)},
    \nonumber\\
    W_Q
    &=\frac{\partial w(t+1)}{\partial Q^h(t)},
    &\,
    W_m
    &=\frac{\partial w(t+1)}{\partial m(t)}.
\end{align}
Differentiating Eq.~\eqref{eq:Phi_prime} and using the fixed\hyph{}point relations
\begin{align}
    Q^h(t+1)=Q^h(t)\equiv Q^h,
    \qquad
    m(t+1)=m(t)\equiv m,
    \qquad
    w(t+1)=2(1-Q^h)m
\end{align}
give
\begin{align}
J_T=
\begin{pmatrix}
\Psi_Q & \Psi_m
\\ \\
\dfrac{W_Q+2m\Psi_Q}{2(1-Q^h)} &
\dfrac{W_m+2m\Psi_m}{2(1-Q^h)}
\end{pmatrix}.
\label{eq:JT_general}
\end{align}

\subsubsection{Single\hyph{}node noise covariance}

For a single\hyph{}node update, introduce the microscopic variables at the next step,
\begin{align}
    X(t+1)&=\mathbf{1}_{s(t+1)=0},
    \qquad
    Y(t+1)=s(t+1),
\end{align}
where $X(t+1)$ is the hidden\hyph{}state indicator and $Y(t+1)$ is the signed state. Let
\begin{align}
    P_0(t+1)\equiv P\bigl(s(t+1)=0\bigr).
\end{align}
The moments of $X(t+1)$ and $Y(t+1)$ are
\begin{align}
    &\langle X(t+1)\rangle = P_0(t+1),
    \qquad
    &\langle Y(t+1)\rangle = w(t+1),
    \\ \nonumber
    &\langle X^2(t+1)\rangle = P_0(t+1),
    \qquad
    &\langle Y^2(t+1)\rangle = 1-P_0(t+1),
    \\ \nonumber
    &\langle X(t+1)Y(t+1)\rangle = 0.
\end{align}
The last equality follows from
\begin{align}
    X(t+1)Y(t+1)
    =\mathbf{1}_{s(t+1)=0}\,s(t+1)
    \equiv0.
\end{align}
The covariance matrix is therefore
\begin{align}
B(t+1)=
\begin{pmatrix}
P_0(t+1)\bigl[1-P_0(t+1)\bigr]
&
-P_0(t+1)w(t+1)
\\ \\
-P_0(t+1)w(t+1)
&
1-P_0(t+1)-w^2(t+1)
\end{pmatrix}.
\label{eq:B_matrix}
\end{align}

The microscopic moments at step $t+1$ determine the macroscopic variables through
\begin{align}
    Q^h(t+1)&=P_0(t+1),
    \qquad
    m(t+1)=\frac{w(t+1)}{2\bigl[1-P_0(t+1)\bigr]}.
    \label{eq:macro_post}
\end{align}
The Jacobian of this transformation is
\begin{align}
J_{\rm mac}
&=
\frac{
\partial\bigl(Q^h(t+1),m(t+1)\bigr)
}{
\partial\bigl(P_0(t+1),w(t+1)\bigr)
}
\nonumber\\ 
&=
\begin{pmatrix}
1 & 0
\\ \\
\dfrac{m(t+1)}{1-P_0(t+1)}
&
\dfrac{1}{2\bigl[1-P_0(t+1)\bigr]}
\end{pmatrix}.
\end{align}
Hence
\begin{align}
    V_{\rm post}
    =J_{\rm mac}B(t+1)J_{\rm mac}^{\top}.
    \label{eq:Vpost_def}
\end{align}

At a fixed point,
\begin{align}
    P_0(t+1)=Q^h,
    \qquad
    w(t+1)=2(1-Q^h)m,
\end{align}
so
\begin{align}
V^*=
\begin{pmatrix}
Q^h(1-Q^h) & 0
\\ \\
0 & \dfrac{1-4m^2}{4(1-Q^h)}
\end{pmatrix}.
\label{eq:Vstar_closed}
\end{align}
Thus, at a fixed point, the single\hyph{}node fluctuations of the hidden fraction and the order parameter are uncorrelated in the $(Q^h,m)$ coordinates. The factor $1-4m^2$ suppresses order\hyph{}parameter fluctuations near full order, $|m|\to1/2$, whereas the factor $(1-Q^h)^{-1}$ enhances order\hyph{}parameter fluctuations as the visible fraction decreases.

\subsubsection{Microscopic increments and batch noise}

Let $\mathbf{z}(t)$ and $\mathbf{z}(t+1)$ denote the pre\hyph{}update and post\hyph{}update single\hyph{}node fluctuation vectors in the $(Q^h,m)$ coordinates. A microscopic update produces
\begin{align}
    \Delta\mathbf{z}(t)
    =\mathbf{z}(t+1)-\mathbf{z}(t).
\end{align}
At a fixed point,
\begin{align}
    \operatorname{Cov}\bigl(\mathbf{z}(t)\bigr)
    =
    \operatorname{Cov}\bigl(\mathbf{z}(t+1)\bigr)
    =V^*.
\end{align}
Under annealed neighborhoods, the post\hyph{}update state is conditionally independent of the pre\hyph{}update state to leading order, so
\begin{align}
\operatorname{Cov}\bigl(\Delta\mathbf{z}(t)\bigr)
=
\operatorname{Cov}\bigl(\mathbf{z}(t+1)\bigr)
+\operatorname{Cov}\bigl(\mathbf{z}(t)\bigr)
-\operatorname{Cov}\bigl(\mathbf{z}(t+1),\mathbf{z}(t)\bigr)
-\operatorname{Cov}\bigl(\mathbf{z}(t),\mathbf{z}(t+1)\bigr)
\simeq 2V^*.
\label{eq:increment_cov_fixed}
\end{align}
Thus, although $\Delta\mathbf{z}(t)$ is a difference, the two variances add, while the minus signs enter only through the vanishing cross\hyph{}covariances.

A batch contains $fN$ microscopic updates, each contributing $N^{-1}\Delta\mathbf{z}(t)$ to the macroscopic variables. Leading\hyph{}order independence among the increments gives
\begin{align}
    \langle\eta_b\eta_b^\top\rangle
    =
    \frac{fN}{N^2}\,2V^*
    =
    \frac{2fV^*}{N}.
\label{eq:batch_noise}
\end{align}

\subsubsection{Steady\hyph{}state covariance and critical divergence}

In the steady state,
\begin{align}
    V=\langle\xi_b\xi_b^\top\rangle
\end{align}
satisfies
\begin{align}
    V
    =
    (I-fA)V(I-fA)^\top
    +
    \frac{2fV^*}{N}.
\label{eq:discrete_lyapunov}
\end{align}
Expanding to first order in $f$ gives
\begin{align}
    (I-fA)V(I-fA)^\top
    =
    V-f(AV+VA^\top)+\mathcal{O}(f^2),
\end{align}
and hence
\begin{align}
    AV+VA^\top=\frac{2V^*}{N}.
\label{eq:lyapunov_SM}
\end{align}
The factor $f$ cancels because deterministic relaxation and stochastic injection accumulate linearly with the number of microscopic updates.

For a $2\times2$ matrix $A$, the solution is
\begin{align}
V=
\frac{
\det A\,V^*
+\operatorname{adj}(A)V^*\operatorname{adj}(A)^\top
}{
N\,\det A\,\operatorname{tr}A
}.
\label{eq:lyapunov_solution}
\end{align}
The first term is regular as $\det A\to0$, whereas the second term contains the singular factor $(\det A)^{-1}$. Using Eq.~\eqref{eq:Vstar_closed}, the leading singular contribution to the order\hyph{}parameter variance $V_{22}=\mathrm{Var}(m)$ is
\begin{align}
V_{22}
&\simeq
\frac{K_V}{N\det A},
\nonumber\\ \nonumber \\
K_V
&=
\frac{
4A_{21}^2Q^h(1-Q^h)^2
+A_{11}^2(1-4m^2)
}{
4(1-Q^h)\operatorname{tr}A
}.
\label{eq:V22_leading}
\end{align}
Along the stable branches considered below, $K_V$ remains finite and nonzero as the critical eigenvalue vanishes. Therefore,
\begin{align}
    V_{22}\propto\frac{1}{\det A}.
\label{eq:V22_det_scaling}
\end{align}
The fluctuation exponent is determined by the scaling of $\det A$ along the occupied fixed\hyph{}point branch.

\subsubsection{Determinant and transition\hyph{}dependent scaling}

The transition\hyph{}dependent scaling follows from the relation between $A$ and the Jacobian of the steady\hyph{}state equations. Following Eqs.~\eqref{eq:visible_equiv} and \eqref{eq:visible_state}, define
\begin{align}
F_1
&=Q^h(t)-Q^h(t+1),
\nonumber\\
F_2
&=2[1-Q^h(t)]m(t)-w(t+1),
\end{align}
where
\begin{align}
w(t+1)
=
2[1-Q^h(t+1)]m(t+1).
\end{align}
At a fixed point,
\begin{align}
Q^h(t+1)=Q^h(t)\equiv Q^h,
\qquad
m(t+1)=m(t)\equiv m,
\end{align}
and the differentials satisfy
\begin{align}
dF_1
&=
dQ^h(t)-dQ^h(t+1),
\nonumber\\
dF_2
&=
-2m\bigl[dQ^h(t)-dQ^h(t+1)\bigr]
+2(1-Q^h)\bigl[dm(t)-dm(t+1)\bigr].
\end{align}
Below, the fixed\hyph{}point variables are written without the time index. Using $A=I-J_T$ gives
\begin{align}
\frac{\partial(F_1,F_2)}
{\partial(Q^h,m)}
=
\begin{pmatrix}
1 & 0
\\
-2m & 2(1-Q^h)
\end{pmatrix}
A,
\label{eq:F_A_relation}
\end{align}
and therefore
\begin{align}
\det
\frac{\partial(F_1,F_2)}
{\partial(Q^h,m)}
=
2(1-Q^h)\det A.
\label{eq:det_identity}
\end{align}
For $Q^h<1$, criticality of the steady\hyph{}state equations is therefore equivalent to $\det A=0$.

\paragraph{Hybrid transition.}
For $p_v<p_v^t$, the ordered fixed point merges with an unstable fixed point at the spinodal point $p_h^{**}$ through a saddle\hyph{}node bifurcation. Let $e=(e_Q,e_m)^\top$ and $u$ be the right and left null vectors of
\begin{align}
\frac{\partial(F_1,F_2)}
{\partial(Q^h,m)}
\end{align}
at $(Q^h_{**},m_{**},p_h^{**})$, where $Q^h_{**}$ and $m_{**}$ are the values at the spinodal point. Expanding
\begin{align}
Q^h
=
Q^h_{**}+\epsilon e_Q+\mathcal{O}(\epsilon^2),
\qquad
m
=
m_{**}+\epsilon e_m+\mathcal{O}(\epsilon^2),
\qquad
p_h
=
p_h^{**}+\delta,
\end{align}
the linear term in $\epsilon$ vanishes because $e$ is the right null vector. Projecting the steady\hyph{}state equations onto $u$, the leading nonvanishing terms are the parameter term $\delta$ and the quadratic term $\epsilon^2$,
\begin{align}
u^\top
\partial_{p_h}
\begin{pmatrix}
F_1\\
F_2
\end{pmatrix}
\delta
+
\frac{1}{2}
u^\top
\partial_{(Q^h,m)}^2
\begin{pmatrix}
F_1\\
F_2
\end{pmatrix}
(e,e)\epsilon^2
=
0.
\label{eq:fold}
\end{align}
Along the ordered branch, $\delta=p_h-p_h^{**}<0$, giving
\begin{align}
\epsilon
\propto
(p_h^{**}-p_h)^{1/2}.
\end{align}
The critical eigenvalue is linear in the displacement along the critical direction,
\begin{align}
\lambda_c
\propto
\epsilon
\propto
(p_h^{**}-p_h)^{1/2}.
\end{align}
Since the noncritical eigenvalue remains finite,
\begin{align}
\det A
\propto
(p_h^{**}-p_h)^{1/2}.
\label{eq:detA_spinodal_scaling}
\end{align}
Substitution into Eq.~\eqref{eq:V22_det_scaling} gives
\begin{align}
\mathrm{Var}(m)
\propto
(p_h^{**}-p_h)^{-1/2}.
\label{eq:C22_leading_spinodal}
\end{align}
This divergence is one\hyph{}sided because only the ordered state approaches the saddle\hyph{}node bifurcation. After the jump to the symmetric state, $\det A$ and $\mathrm{Var}(m)$ remain finite.

\paragraph{Continuous transition.}
For $p_v>p_v^t$, the ordered fixed point merges continuously with the symmetric fixed point at the continuous transition point $p_h^{c}$. Let $e=(e_Q,e_m)^\top$ and $u$ be the right and left null vectors of
\begin{align}
\frac{\partial(F_1,F_2)}
{\partial(Q^h,m)}
\end{align}
at $(Q_0^h,0,p_h^{c})$. At the symmetric fixed point, the $\mathbb{Z}_2$ transformation $m\to-m$ leaves $Q^h$ invariant. The hidden\hyph{}fraction sector is even in the displacement $\epsilon$, whereas the order\hyph{}parameter sector is odd. The critical direction therefore lies in the order\hyph{}parameter sector, so $e_Q=0$, and the fixed\hyph{}point expansion takes the symmetry\hyph{}compatible form
\begin{align}
Q^h
=
Q_0^h+\mathcal{O}(\epsilon^2),
\qquad
m
=
\epsilon e_m+\mathcal{O}(\epsilon^3),
\qquad
p_h
=
p_h^{c}+\delta.
\end{align}
Projecting the steady\hyph{}state equations onto the left null vector $u$ selects the order\hyph{}parameter sector. The resulting scalar equation must be odd under $\epsilon\to-\epsilon$,
\begin{align}
u^\top
\begin{pmatrix}
F_1\\
F_2
\end{pmatrix}
(-\epsilon,\delta)
=
-
u^\top
\begin{pmatrix}
F_1\\
F_2
\end{pmatrix}
(\epsilon,\delta).
\end{align}
Terms even in $\epsilon$, including $\delta$ and $\epsilon^2$, are therefore forbidden. The null\hyph{}vector condition removes the term linear in $\epsilon$ at $\delta=0$. Away from the transition, the leading linear contribution is $\delta\epsilon$, while the leading nonlinear contribution allowed by the $\mathbb{Z}_2$ symmetry is $\epsilon^3$. The projected steady\hyph{}state equation becomes
\begin{align}
u^\top
\partial_{(Q^h,m)p_h}^2
\begin{pmatrix}
F_1\\
F_2
\end{pmatrix}
(e)\,\delta\epsilon
+
\frac{1}{6}
u^\top
\partial_{(Q^h,m)}^3
\begin{pmatrix}
F_1\\
F_2
\end{pmatrix}
(e,e,e)\epsilon^3
+
\mathcal{O}\!\left(
\delta^2\epsilon,
\delta\epsilon^3,
\epsilon^5
\right)
=
0.
\label{eq:pitchfork}
\end{align}

On the ordered side, $\epsilon\neq0$ and $\delta=p_h-p_h^{c}<0$, giving
\begin{align}
\epsilon^2
\propto
p_h^{c}-p_h.
\end{align}
The critical eigenvalue is quadratic in the displacement along the critical direction,
\begin{align}
\lambda_c
\propto
\epsilon^2
\propto
p_h^{c}-p_h,
\end{align}
and hence
\begin{align}
\det A
\propto
p_h^{c}-p_h,
\qquad
p_h<p_h^{c}.
\end{align}

On the symmetric side, $\epsilon=0$. The critical eigenvalue is then controlled by the linear coefficient of the projected equation and scales as
\begin{align}
\lambda_c
\propto
\delta
=
p_h-p_h^{c}.
\end{align}
Since the noncritical eigenvalue remains finite,
\begin{align}
\det A
\propto
p_h-p_h^{c},
\qquad
p_h>p_h^{c}.
\end{align}
Combining the two sides gives
\begin{align}
\det A
\propto
|p_h-p_h^{c}|.
\label{eq:detA_continuous_scaling}
\end{align}
Substitution into Eq.~\eqref{eq:V22_det_scaling} therefore gives
\begin{align}
\mathrm{Var}(m)
\propto
|p_h-p_h^{c}|^{-1}.
\label{eq:C22_leading}
\end{align}
The continuous transition thus exhibits a two\hyph{}sided divergence.

Both transitions obey $\mathrm{Var}(m)\propto(\det A)^{-1}$ but have different singular behavior. At the spinodal point, $\det A\propto(p_h^{**}-p_h)^{1/2}$ only on the ordered side, giving a one\hyph{}sided variance exponent $1/2$. At the continuous transition point, $\det A\propto|p_h-p_h^{c}|$ on both sides, giving a two\hyph{}sided variance exponent $1$.

\section{External fields}
\label{app:external_fields}

Fields are applied by reweighting the outcome probabilities of each update,
\begin{align}
P_{\text{eff}}(s)=\frac{P_{\text{bare}}(s)e^{h_s}}
{\sum_{s'}P_{\text{bare}}(s')e^{h_{s'}}},
\qquad
h_{s=0}\equiv h_0,
\qquad
h_{s=\pm1}\equiv\pm h.
\label{eq:field_reweight}
\end{align}
Here, $h$ denotes the field conjugate to the visible opinion, and $h_0$ denotes the field conjugate to concealment. In stage (ii) with a $+1$ majority, the bare probabilities $(1-p_h,\,0,\,p_h)$ for $s(t+1)=(+1,-1,0)$ become
\begin{align}
P_{\text{eff}}(+1) &= \frac{(1-p_h)e^h}{Z_+},
\qquad
P_{\text{eff}}(-1) = 0,
\qquad
P_{\text{eff}}(0) = \frac{p_h e^{h_0}}{Z_+},
\qquad {\rm where} \quad Z_+= (1-p_h)e^h+p_h e^{h_0}.
\end{align}
In stage (ii) with a $-1$ majority, the roles of $\pm1$ are exchanged,
\begin{align}
P_{\text{eff}}(+1) &= 0,
\qquad
P_{\text{eff}}(-1) = \frac{(1-p_h)e^{-h}}{Z_-},
\qquad
P_{\text{eff}}(0) = \frac{p_h e^{h_0}}{Z_-},
\qquad {\rm where}
Z_- &= (1-p_h)e^{-h}+p_h e^{h_0}.
\end{align}
In stage (iii) (free choice),
\begin{align}
P_{\text{eff}}(+1) &= \frac{(p_v/2)e^h}{Z_0},
\qquad
P_{\text{eff}}(-1) = \frac{(p_v/2)e^{-h}}{Z_0},
\qquad
P_{\text{eff}}(0) = \frac{(1-p_v)e^{h_0}}{Z_0},
\qquad {\rm where}
Z_0 &= \frac{p_v}{2}\left(e^h+e^{-h}\right)+(1-p_v)e^{h_0}.
\end{align}

The reweighted probabilities replace the bare probabilities in Eqs.~\eqref{eq:psi_h_full}\babelhyphen{--}\eqref{eq:sigma_h} of Sec.~\ref{app:coefficients}. At zero opinion field, stage (iii) assigns $+1$ and $-1$ with equal probability. A nonzero opinion field $h$ breaks this equality at first order and adds a stage\hyph{}(iii) contribution to the response of $m$. The opinion field $h$ also breaks the $\mathbb{Z}_2$ symmetry, so $C_0$, $C_2$, and $C_4$ need not vanish. By contrast, the concealment field $h_0$ preserves the $\mathbb{Z}_2$ symmetry. The finite\hyph{}field fixed points are therefore obtained by direct numerical solution of the reweighted equations.

\subsection{Linear response}
\label{app:response_general}

The linear response to the external fields $(h_0,h)$ follows from the field dependence of the steady\hyph{}state equations. For the one\hyph{}step map, collect the field derivatives of the output quantities in
\begin{align}
S=
\begin{pmatrix}
\partial_{h_0}Q^h(t+1) & \partial_h Q^h(t+1)
\\ \\
\partial_{h_0}w(t+1) & \partial_h w(t+1)
\end{pmatrix}_{h_0=h=0},
\label{eq:S_def}
\end{align}
where the derivatives are taken at fixed input $(Q^h(t),m(t))$.

The steady\hyph{}state equations are
\begin{align}
F_1
&=Q^h(t)-Q^h(t+1)=0,
\nonumber\\
F_2
&=2[1-Q^h(t)]m(t)-w(t+1)=0.
\end{align}
At a fixed point, the time indices are suppressed. Differentiating the steady\hyph{}state equations with respect to $(h_0,h)$ gives
\begin{align}
\frac{\partial(Q^h,m)}
{\partial(h_0,h)}
=
\frac{
\operatorname{adj}\!\left[
\dfrac{\partial(F_1,F_2)}
{\partial(Q^h,m)}
\right]S
}{
\det
\dfrac{\partial(F_1,F_2)}
{\partial(Q^h,m)}
}.
\label{eq:response_matrix}
\end{align}
Using Eq.~\eqref{eq:det_identity}, the susceptibility
\begin{align}
\chi_{\mathrm{MF}}
\equiv
\left.\frac{\partial m}{\partial h}\right|_{h_0=h=0}
\end{align}
has the critical form
\begin{align}
\chi_{\mathrm{MF}}
\simeq
\frac{K_\chi}{\det A},
\qquad
K_\chi
=
\frac{
\left[
\operatorname{adj}\!\left(
\dfrac{\partial(F_1,F_2)}
{\partial(Q^h,m)}
\right)S
\right]_{22}
}{
2(1-Q^h)
}.
\label{eq:chi_detA}
\end{align}
For the transitions considered below, $K_\chi$ remains finite and nonzero as the critical eigenvalue vanishes. The divergence of $\chi_{\mathrm{MF}}$ is therefore controlled by the same factor $\det A$ as the divergence of $\mathrm{Var}(m)$.

\paragraph{Hybrid transition.}
For $p_v<p_v^t$, the ordered state approaches the spinodal point $p_h^{**}$, where the ordered state loses stability. Using Eq.~\eqref{eq:detA_spinodal_scaling},
\begin{align}
\det A
\propto
(p_h^{**}-p_h)^{1/2},
\qquad
p_h<p_h^{**},
\end{align}
and hence
\begin{align}
\chi_{\mathrm{MF}}
\propto
(p_h^{**}-p_h)^{-1/2}.
\label{eq:chi_spinodal}
\end{align}
The divergence is one\hyph{}sided: $\chi_{\mathrm{MF}}$ diverges on approaching the spinodal point from the ordered state, and the divergence terminates at the jump. In the symmetric state reached after the jump, $\det A$ and $\chi_{\mathrm{MF}}$ remain finite.

\paragraph{Continuous transition.}
For $p_v>p_v^t$, the ordered state continuously approaches the symmetric state at the continuous transition point $p_h^{c}$. Using Eq.~\eqref{eq:detA_continuous_scaling},
\begin{align}
\det A
\propto
|p_h-p_h^{c}|,
\end{align}
on both sides of the transition, and hence
\begin{align}
\chi_{\mathrm{MF}}
\propto
|p_h-p_h^{c}|^{-1}.
\label{eq:chi_continuous}
\end{align}
The divergence is therefore two\hyph{}sided.

The susceptibility and the order\hyph{}parameter variance obey the common critical scaling
\begin{align}
\chi_{\mathrm{MF}}\propto\frac{1}{\det A},
\qquad
\mathrm{Var}(m)\propto\frac{1}{\det A}.
\end{align}
At the spinodal point, $\det A\propto(p_h^{**}-p_h)^{1/2}$ on the ordered side, and the susceptibility exponent is $\gamma=1/2$. At the continuous transition point, $\det A\propto|p_h-p_h^{c}|$ on both sides, and the susceptibility exponent is $\gamma=1$.

The common singular factor $\det A$ cancels in the amplitude ratio,
\begin{align}
\frac{N\,\mathrm{Var}(m)}{\chi_{\mathrm{MF}}}
\longrightarrow
\frac{K_V}{K_\chi}.
\label{eq:amplitude_ratio}
\end{align}
The ratio therefore remains finite when either transition is approached along a critical side. Detailed balance is absent, so the equilibrium fluctuation\babelhyphen{--}dissipation relation does not fix the limiting value of the ratio~\cite{marconi2008response}.

\section{Entropy flux formula}
\label{app:entropy_flux}

Let $P(s)$ be the steady\hyph{}state probability of the single\hyph{}node state $s\in\{+1,-1,0\}$, obtained from the mean\hyph{}field fixed point $(Q^h,m)$ and Eq.~\eqref{eq:Qpm_x} as
\begin{align}
    P(0) = Q^h,
    \qquad
    P(\pm1) = Q^{\pm} = \frac{1-Q^h}{2}\,(1\pm2m).
    \label{eq:Ps_def}
\end{align}
The neighborhood composition $\mathbf{n}=(v_+,v_-,n_h)$, subject to $v_++v_-+n_h=k$, is drawn from the multinomial distribution
\begin{align}
    P(\mathbf{n})
    =
    \frac{k!}{v_+!\,v_-!\,n_h!}
    P(+1)^{v_+}P(-1)^{v_-}P(0)^{n_h}.
    \label{eq:env_dist}
\end{align}
This distribution is exact for the annealed neighborhoods of the main text, in which the $k$ neighbors are redrawn uniformly at random at every update.

Let $v=v_++v_-$ denote the number of visible neighbors. The update rule of the main text assigns the post\hyph{}update probabilities
\begin{align}
P\bigl(s(t+1)\mid\mathbf{n}\bigr)
=
\begin{cases}
\bigl(1-p_h,\,0,\,p_h\bigr),
& \text{stage (ii)},\quad v_+>\ceil*{v/2},
\\
\bigl(0,\,1-p_h,\,p_h\bigr),
& \text{stage (ii)},\quad v_->\ceil*{v/2},
\\
\bigl(\tfrac{p_v}{2},\,\tfrac{p_v}{2},\,1-p_v\bigr),
& \text{stage (iii),\quad free choice},
\end{cases}
\label{eq:outcome_prob}
\end{align}
for $s(t+1)=(+1,-1,0)$. Stage (ii) applies when $\max(v_+,v_-)>\ceil*{v/2}$, and stage (iii) applies otherwise.

The entropy flux of Eq.~\eqref{eq:entropy_flux} is resolved by neighborhood composition. Each composition $\mathbf{n}$ defines a separate channel, and the total flux is the weighted sum
\begin{align}
    \Sent
    &=
    \sum_{\mathbf{n}}P(\mathbf{n})\,\sigma(\mathbf{n}),
    \nonumber\\
    \sigma(\mathbf{n})
    &=
    \sum_{s(t)}P\bigl(s(t)\bigr)
    \sum_{s(t+1)\neq s(t)}
    P\bigl(s(t+1)\mid\mathbf{n}\bigr)
    \ln
    \frac{
    P\bigl(s(t+1)\mid\mathbf{n}\bigr)
    }{
    P\bigl(s(t)\bigr)
    }.
    \label{eq:entropy_flux_SM}
\end{align}
Here, $\sigma(\mathbf{n})$ is the flux carried by channel $\mathbf{n}$. The first sum runs over the pre\hyph{}update state $s(t)$, and the second sum runs over the post\hyph{}update state $s(t+1)$. The restriction $s(t+1)\neq s(t)$ excludes updates without a state change.

The numerator $P\bigl(s(t+1)\mid\mathbf{n}\bigr)$ is the probability that the node adopts state $s(t+1)$, given the neighborhood composition $\mathbf{n}$. The update rule overwrites the pre\hyph{}update state, so this probability does not depend explicitly on $s(t)$. The denominator $P\bigl(s(t)\bigr)$ is the steady\hyph{}state probability of the pre\hyph{}update state. Under annealed neighborhoods, the node state is independent of the sampled neighborhood composition, so every channel uses the steady\hyph{}state probability $P(s)$ of Eq.~\eqref{eq:Ps_def}. Equation~\eqref{eq:entropy_flux_SM} defines the neighborhood\hyph{}resolved observable used in this work. Unlike the standard entropy flux, it does not involve reverse transition rates, and we do not identify it with the thermodynamic entropy production. Related studies of entropy production in opinion models provide the broader context~\cite{crochik2005entropy,tome2012entropy,hawthorne2023thermodynamics,oliveira2024entropy}.

At $(p_h,p_v)=(0,1)$, the model reduces to the conventional majority\hyph{}rule voter model and reaches an absorbing consensus. For the $+1$ consensus,
\begin{align}
    Q^h=0,
    \qquad
    m=\tfrac12,
    \qquad
    P(+1)=1.
    \nonumber
\end{align}
The only neighborhood composition with nonzero probability is $v_+=k$, for which stage (ii) gives
\begin{align}
    P\bigl(s(t+1)=+1\mid\mathbf{n}\bigr)=1.
    \nonumber
\end{align}
No state\hyph{}changing update occurs, and therefore $\Sent=0$.

At $p_h=1$, stage (ii) always assigns the hidden state, whereas stage (iii) assigns a visible state with probability $p_v$. The steady\hyph{}state visible fraction therefore satisfies
\begin{align}
    1-Q^h=p_v\mathcal{P}_{\mathrm{free}},
    \nonumber
\end{align}
where $\mathcal{P}_{\mathrm{free}}$ is the steady\hyph{}state probability that the sampled neighborhood satisfies the free\hyph{}choice condition of stage (iii). Every state\hyph{}changing update involves a visible state, so the steady\hyph{}state visible fraction controls the rate of state\hyph{}changing updates. The high\hyph{}concealment regime ($p_v=0.1$) therefore has a smaller $\Sent$ than the low\hyph{}concealment regime ($p_v=0.4$), although both regimes are in the symmetric state with $m=0$.

For $k=8$, the sum in Eq.~\eqref{eq:entropy_flux_SM} is evaluated exactly over all
\begin{align}
    \binom{k+2}{2}=45
    \nonumber
\end{align}
neighborhood compositions at the mean\hyph{}field fixed points used in Figs.~\ref{fig:phase_diagram} and \ref{fig:entropy}.

\subsection{Single\hyph{}node trajectories in the symmetric state}
\label{app:trajectory}

\begin{figure}[h]
\includegraphics[width=\textwidth]{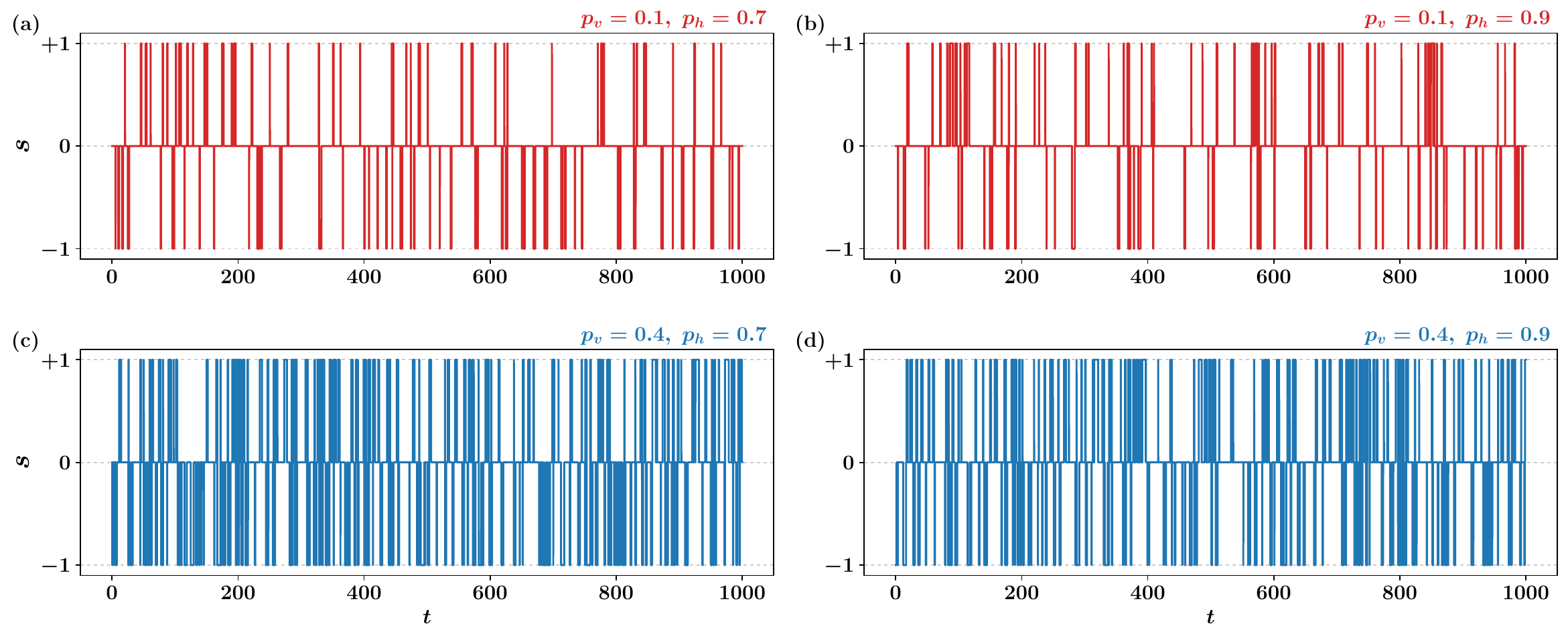}
\caption{State evolution $s_i(t)$ of a representative node $i$ in MC simulations for (a) $(p_h,p_v)=(0.7,0.1)$, (b) $(0.9,0.1)$, (c) $(0.7,0.4)$, and (d) $(0.9,0.4)$. The time series $s_i(t)$ are shown over the last $10^3$ MC sweeps after reaching the steady state. For $p_v=0.1$ [(a) and (b)], the node switches only infrequently from the hidden state $s=0$. In contrast, for $p_v=0.4$ [(c) and (d)], the node switches frequently between the hidden and visible states. Here, $N=16\times10^4$.
}
\label{fig:trajectory}
\end{figure}

Equation~\eqref{eq:entropy_flux_SM} counts only state\hyph{}changing updates, $s(t+1)\neq s(t)$. To show these updates directly, we record the state $s_i(t)$ of a single node ($i=7$) in MC simulations with $N=16\times10^4$ over the last $10^3$ MC sweeps in the steady state. Figure~\ref{fig:trajectory} shows $s_7(t)$ for $p_v=0.1$ and $0.4$ at $p_h=0.7$ and $0.9$. All four parameter sets lie above the transition point, so the system is in the symmetric state with $m=0$.

For $p_v=0.1$, the node remains in the hidden state $s=0$ most of the time and enters the visible states $s=\pm1$ only briefly. For $p_v=0.4$, the node switches between the hidden and visible states much more frequently. In both regimes, the node visits $s=+1$ and $s=-1$ with similar frequency, consistent with $m=0$. The two regimes therefore share the same order parameter but differ in the rate of state\hyph{}changing updates. This difference is the microscopic origin of the smaller $\Sent$ in the high\hyph{}concealment regime and the larger $\Sent$ in the low\hyph{}concealment regime (Fig.~\ref{fig:entropy}).

\section{Monte Carlo protocol}
\label{app:mc_protocol}

Updates are applied in batches of $\sqrt{N}$ nodes. Each batch selects $\sqrt{N}$ distinct nodes at random. For each selected node, $k$ neighbors are drawn without replacement with probability proportional to degree, which reduces to uniform sampling for the fixed degree used here. The batch states are computed from the current configuration and updated simultaneously. One sweep contains $N/\sqrt{N}=\sqrt{N}$ batches and updates every node once on average. Observables are recorded after each batch, giving $\approx\sqrt{N}$ samples per sweep.

Each run contains $n_{\text{iter}}={2.56 \times 10^{10}}/{N}$ sweeps, and the first $\sqrt{n_{\text{iter}}}$ sweeps are discarded as thermalization. At $N=16\times10^4$, $n_{\text{iter}}=1.6\times10^{5}$. The discarded window is long enough to reach the steady state at every parameter point. At $N=16\times10^4$, the retained samples number $\simeq6.4\times10^{7}$ for each of $m$ and $Q^h$. The samples provide estimates of $\langle m \rangle$ through $\langle m^4 \rangle$ and $\langle Q^h\rangle$. The long averaging window provides the statistical accuracy. Each parameter point therefore uses a single run.

For each $p_v\in\{0.1,0.4,0.5\}$, $p_h$ is varied over $[0,1]$ in steps of 0.05. 
Near a transition, grid refinement is performed by first locating the maximum of $\mathrm{Var}(m)=\langle m^2\rangle-\langle m\rangle^2$ on the coarse grid. The bracketing interval is then divided into ten equal subintervals. If the maximum lies away from the center of the refined window, the window is re\hyph{}centered accordingly. The values $p_v=0.1$ and $0.4$ represent the high\hyph{} and low\hyph{}concealment regimes, respectively. The additional value $p_v=0.5$ is included to verify the low\hyph{}concealment behavior at a second parameter value.

\renewcommand{\bibsection}{\section*{Supplemental References}}
\end{bibunit}

\end{document}